\documentclass[12pt]{article}
\usepackage{graphicx}
\usepackage{amsfonts,amsmath}
\usepackage[mathscr]{eucal}
\usepackage{amssymb}
\usepackage{amsthm}
\usepackage{bbold}
\theoremstyle{plain}

\def\theequation{\arabic{section}.\arabic{equation}}
\newcommand{\be}{\begin{eqnarray}}
\newcommand{\ee}{\end{eqnarray}}
\newcommand{\bc}{\begin{center}}
\newcommand{\ec}{\end{center}}
\newcommand{\nn}{\nonumber \\}
\newcommand{\lb}{\label}
\newcommand{\p}[1]{(\ref{#1})}
\newcommand{\vecg}[1]{\mbox{\boldmath $#1$}}
\renewcommand{\u}{\underline}

\begin{document}

\begin{titlepage}

\vspace*{0.2cm}

\renewcommand{\thefootnote}{\star}
\begin{center}

{\LARGE\bf  An 8-dimensional Taub-NUT-like hyper-K\"ahler metric in harmonic superspace formalism}

\vspace{2cm}

{\Large A.V. Smilga} \\

\vspace{0.5cm}

{\it SUBATECH, Universit\'e de
Nantes,  4 rue Alfred Kastler, BP 20722, Nantes  44307, France. }

\end{center}
\vspace{0.2cm} \vskip 0.6truecm \nopagebreak

   \begin{abstract}
\noindent  
Using the harmonic superspace  formalism, we find the metric of a certain 8-dimensional manifold. This manifold is not compact and represents an  8-dimensional generalization of the Taub-NUT manifold. Our conjecture is that the metric that we derived is equivalent to the known metric possessing a discrete $Z_2$ isometry, which may be obtained from the metric describing the dynamics  of four BPS monopoles by Hamiltonian reduction.
   \end{abstract}

\end{titlepage}

\setcounter{footnote}{0}

\setcounter{equation}0
\section{Introduction}

Hyper-K\"ahler (HK) manifolds admit a triple of covariantly constant quaternionic complex structures,
\be 
\lb{3-structures}
\nabla_M I_{NL}^{p = 1,2,3} \ =\ 0,  \qquad (I^p)_M{}^N  (I^q)_N{}^L \ =\ -\delta^{pq} \delta_M^L + \varepsilon^{pqr} (I^r)_M{}^L  \,.
 \ee
 Their dimension is an integer multiple of 4,
 \be
 \lb{D=4n}
  D \ =\ 4n \,.
  \ee
  It is well known that the supersymmetric sigma models with HK target spaces enjoy extended supersymmetries. This is true for field-theory models \cite{AG-F} and also for supersymmetric quantum mechanical (SQM) models where the variables and supervariables depend only on time, but not on spatial coordinates.
  
  Such HK models including $4n$ real  ({\bf 1}, {\bf 2}, {\bf 1}) supermultiplets \footnote{We use the notation introduced in \cite{Toppan}:  the first number counts the dynamical real bosonic variables, the second number counts the dynamical real fermionic variables and the third number counts the bosonic auxiliary variables.} possess ${\cal N} = 8$ supersymmetry, and the models including $2n$ chiral  ({\bf 2}, {\bf 2}, {\bf 0}) supermultiplets possess ${\cal N} = 4$ supersymmetry. In this article, we will only discuss the second type of models.
  
  The ${\cal N} = 4$ SQM superfields depend on time, on $\theta_{j=1,2}$ and on $\bar \theta^j$. A generic such superfield,
  \be
  \lb{Phi-gen}
  \Phi(t, \theta_j, \bar \theta^j) \  = \ \phi(t) + \theta_j \psi^j(t) + \ldots,
   \ee
   involves 16 complex component variables. And  ${\cal N} = 8$ superfields depending
   on $t, \theta_{j\alpha}, \bar \theta^{j\alpha}$ involve in the expansion 256 complex components.  
   But this description is not convenient. In particular, it is difficult in the ordinary language to find the constraints to be imposed on a generic superfield to make the constrained superfield an irreducible representation of the supersymmetry algebra.
   
   This can be done, however, in the {\it harmonic superspace} (HSS) formalism
   \cite{HSS}. In \cite{L+4-HK}, this formalism was applied to describe  the 
   ${\cal N}=8$ supersymmetric sigma models living on HK target spaces.  The ${\cal N} = 4$ models living on HK manifolds were described in 
   \cite{DI}.\footnote{The formalism developped in Ref. \cite{DI} also describes  the so-called HKT manifolds, where the complex structures are covariantly constant with respect not to the ordinary Levi-Civita connections, but to certain special {\it Bismuth connections} including nonzero torsions. But we will be interested only in the HK manifolds in this paper.}
   Choosing a harmonic prepotential ${\cal L}^{+4}$, it is possible to find the corresponding HK metric.
   
   However, in contrast to K\"ahler manifolds, where the metric is derived from the K\"ahler potential by simple differentialtion, $h_{j\bar k} = \partial_j \partial_{\bar k} {\cal K}(z^n, \bar z^n)$,  the hyper-K\"ahler problem is much more difficult. To derive the metric from ${\cal L}^{+4}$, one has to solve a set of rather complicated harmonic differential equations. One can always do so
 perturbatively in the neighbourhood of a given point on the manifold, but only in rare cases one can find an analytic solution to these equations. In \cite{L+4-HK}, this was explicitly done only for two 4-dimensional HK manifolds: the Taub-NUT manifold and the Eguchi-Hanson manifold.
 
 In this paper, we explicitly derive in this way a particular 8-dimensional HK metric. This metric is not compact and represents a generalization of the Taub-NUT metric.

\section{One-dimensional harmonic superspace.}
\setcounter{equation}0

With the main purpose to establish notations, we present here the very essentials of the harmonic superspace formalism. For more  details, which are necessary for understanding, the reader is addressed to the monography \cite{HSS}, to the paper \cite{IL}, where the HSS technique, invented originally to deal with extended supersymmetric field theories, was adapted for SQM systems, and to Chap. 7.4 and Chap. 12 of recent \cite{glasses}. 

The harmonic superspace technique  capitalizes on the presence of the $R$-symmetry 
that rotates complex supercharges and the associated fermion variables. For ${\cal N} = 4$ systems, this symmetry is $U(2)$ and its relevant for the following part is $SU(2)$.  The basic idea is to {\it extend} the ${\cal N} = 4$ superspace $(t; \theta_j, \bar \theta^j)$ by adding  extra commuting coordinates $u^{+j}$ parameterizing this $SU(2)$ group. $u^{+j}$ is a complex spinor of unit length:
 \be
\lb{norm-u}
u^{+j} \overline{u^{+j}} \ \stackrel {\rm def}= \ u^{+j}u^-_j \ = \ 1 \,.
 \ee
 The index $j$ is lowered and lifted by $\varepsilon_{jk} = -\varepsilon^{jk}$ with the convention $\varepsilon_{12} = 1$. The matrix
    \be
    \lb{u+u-matr}
    U \ =\ \left( \begin{array}{cc} u^+_1   &  u^-_1 \\ u^+_2 & u^-_2 \end{array} \right)
    \ee
    belongs to $SU(2)$ so that 
 the property
  \be
  \lb{uu-uu}
  u^+_j u^-_k - u^-_j u^+_k \ =\ \varepsilon_{jk}
   \ee
   holds.
   
 Introduce now the odd coordinates $\theta^\pm = \theta^j u^\pm_j, \ \ \ \bar \theta^\pm = \bar \theta^j u^\pm_j$. They are  $SU(2)$ invariants.
  The harmonic superspace is parameterized by the coordinates 
   \be
\lb{full-harm} 
 t; \ \theta^+, \bar \theta^+, \theta^-, \bar \theta^-; \ u^\pm_j \, .
 \ee
 One can define a subspace of this superspace (called {\it Grasmann-analytic} superspace) parameterized by $(t_A;\ \theta^+, 
\bar \theta^+; \ u^\pm_j )$, 
where 
\be
\lb{anal-time}
t_A \ =\ t + i(\theta^+ \bar \theta^- + \theta^- \bar \theta^+) 
 \ee 
is the so-called {\it analytic time}. The point is that the Grassmann-analytic superspace is invariant under supersymmetry transformations.\footnote{The analogy with familiar chiral superspaces is straightforward.}. 
It is also invariant under the  special {\it pseudoconjugation}  operation---a combination of the ordinary complex conjugation and the following transformation of the harmonics: $u^+_j  \to  u^-_j, \ u^-_j  \to  - u^+_j$. The pseudoconjugation acts as 
\begin{equation}\label{harm-tilde}
\widetilde{\theta^\pm}=\bar\theta^\pm\,,\qquad \widetilde{\bar\theta^\pm}=-\theta^\pm\,,\qquad
\widetilde{u^\pm_j}=u^\pm{}^j\,.
\end{equation}
 
 Introduce the supersymmetric harmonic  derivatives:
 \be
\lb{D-harm}
D^0 &=&  \partial^0 \  +  \ \theta^+ \frac {\partial}{\partial \theta^+ }
+  \bar\theta^+ \frac {\partial}{\partial \bar\theta^+} -  \theta^- \frac {\partial}{\partial \theta^- } -
\bar \theta^- \frac {\partial}{\partial \bar \theta^- }\,, \nn
D^{++} &=& \partial^{++}  +\
\theta^+ \frac {\partial}{\partial \theta^- } +  \bar \theta^+ \frac {\partial}{\partial \bar\theta^- } \,,
\nn
D^{--} &=&  \partial^{--} \ + \
\theta^- \frac {\partial}{\partial \theta^+ } +  \bar \theta^- \frac {\partial}{\partial \bar\theta^+ } \,,
\ee
where 
\be
\lb{d-harm}
\partial^0 &=&  u_j^+ \frac {\partial}{\partial u_j^+ } -  u_j^- \frac {\partial}{\partial u_j^- }\,, \nn
\partial^{++} &=& u^+_j \frac {\partial}{\partial u_j^-}\,, \nn
 \partial^{--} &=& u^-_j \frac {\partial}{\partial u_j^+} \,.
  \ee
Both \p{D-harm} and \p{d-harm} satisfy the $su(2)$ algebra.
The operator $D^0$ is called the {\it harmonic charge}. All the harmonic superfields that we are going to deal with have a definite integer value of this charge. The operator $D^{++}$ acting on a superfield of harmonic charge $q$ produces a superfield of  charge $q+2$.  The operator $D^{--}$ acting on a  superfield of charge $q$ produces a  superfield of charge $q - 2$.

Consider a Grassmann-analytic superfield of harmonic charge $+1$:
\be
\lb{q+}
q^+(t_A; \theta^+, \bar \theta^+; u)  = f^+(t_A, u) + \theta^+ \chi(t_A, u) + 
\bar \theta^+ \kappa(t_A, u) + \theta^+ \bar \theta^+ A^-(t_A, u), \qquad
 \ee
where $ f^+(t_A, u)$ and $A^-(t_A, u)$ are Grassmann-even variables of \index{harmonic charge}  harmonic charge $+1$ and $-1$, respectively, and $\chi(t_A, u)$ and $ \kappa(t_A, u)$
are Grassmann-odd variables of zero \index{harmonic charge}  harmonic charge.

The absence of the dependence of $q^+$ on $\theta^-$ and $\bar \theta^-$ makes the component expansion short. However, the components now depend not only on time, but represent infinite series in $u^+$ and $u^-$. 
A remarkable fact, however, is that one can eliminate almost all the terms in these series by imposing a constraint
  \be
\lb{lin-cons}
  D^{++} q^+ \ =\ 0 \, .
  \ee
A general solution to the constraint \p{lin-cons} is
  \be
\lb{q+sol}
q^+(\zeta, u) \ =\ f^j(t_A)u^+_j  + 
\theta^+ \chi(t_A) + 
\bar \theta^+ \kappa(t_A)  - 2i \theta^+ \bar \theta^+\dot{f}^j(t_A) u^-_j\, ,
 \ee
  We are left in this case with only two complex bosonic variables $f^j(t_A)$ 
and two complex fermionic variables $\chi(t_A)$ and $\kappa(t_A)$. 
 
 To describe a sigma model living on a $4n$-dimensional HK manifold, we take $2n$ superfields \p{q+} and impose, instead of the constraints \p{lin-cons}, a nontrivial nonlinear constraint \cite{DI}
  \be
\lb{nonlin-cons}
D^{++} q^{+a} \ =\ {\cal L}^{+3a}  \equiv \Omega^{ab} \, \frac {\partial {\cal L}^{+4}(q^{+c}, u) }{\partial q^{+b}}\, ,
  \ee
  where ${\cal L}^{+4}$ is an arbitrary function of its arguments carrying harmonic charge $+4$ and
  $\Omega^{ab}$ is the symplectic matrix, which can be chosen as 
\be 
\lb{Omega}
  \Omega^{ab} \ =\ -{\rm diag} (\varepsilon, \ldots, \varepsilon)\,.
  \ee
  The symplectic index $a$ will be lifted and lowered by multiplication over $\Omega^{ab}$ and $\Omega_{ab} = - \Omega^{ab}$.
 
 In addition, we impose on $q^{+a}$ the pseudoreality constraint 
  \be
  \lb{harm-pseudoreal}
  \widetilde{q^{+a}} = - q^+_a \,.
  \ee 
 
 Under these conditions, the invariant action 
 \be
 \lb{action}
 S  \ =\  - \frac 18  \Omega_{ab} \int dt du \, d\theta^+ d \bar \theta^+  d\theta^- d \bar \theta^-\, q^{+a} q^{-b},  
  \ee
  where $q^{-a} = D^{--} q^{+a}$, 
  describes the HK geometry \footnote{For a generic ${\cal L}^{+3a}$ in \p{nonlin-cons}, we would arrive to HKT geometry, but this is outside the scope of this paper.} (the coefficient $-1/8$ is a convenient convention). It enjoys  ${\cal N} = 4$ supersymmetry by construction. The expression \p{action} involves, besides the familiar time and $\theta$ integration, also the untegral over the $SU(2)$ group that the harmonics $u^{+j}$ parameterize. This integral can be done using the properties \p{norm-u} and \p{uu-uu}. If we normalize the group volume $\int \! du$ to 1, we derive
  \be
   \lb{harm-int}
  && \int u^+_j u^-_k  \, du \ = \ \frac 12 \varepsilon_{jk}, \nn
  &&\int u^+_j u^+_k u^-_l u^-_p \, du \ =\   
    \frac 16( \varepsilon_{jl} \varepsilon_{kp} +  
    \varepsilon_{jp} \varepsilon_{kl}) \, , \qquad {\rm etc.}
    \ee
   The integrals of the monoms including  inequal numbers of $u^+$ and $u^-$ vanish.
  
  To make contact with geometry, we have to express the action \p{action} in components. To begin with, we substitute the expansion \p{q+} into the constraints \p{nonlin-cons}. We derive for the components
   \be
\lb{cons-comp}
\partial^{++} f^{+ a} &=& \Omega^{ab} \, \frac{\partial {\cal L}^{+4}(f^{+c}, u)}{\partial f^{+b}} \,,  \\
{\cal D}^{++} \chi^a &=& {\cal D}^{++} \kappa^a = 0\,, \lb{ferm} \\
{\cal D}^{++} A^{- a} &=&  -2i \dot{f}^{+ a} + \ \Omega^{ab} \frac
{\partial^3 {\cal L}^{+4}(f^{+e}, u)}{\partial f^{+b} \partial f^{+c}  \partial f^{+d}}\,\kappa^c\chi^d\,,
\lb{A-}
\ee
where the action of the covariant harmonic derivative ${\cal D}^{++}$ 
on any symplectic vector $X^a$ is defined as 
       \be
       \lb{D++def-HKT}
{\cal D}^{++} X^a \ =\     \partial^{++} X^a   -  \,\Omega^{ab} \frac {\partial^2 {\cal L}^{+4}}{\partial f^{+b}\partial f^{+c}} \, X^c \,, \nn
{\cal D}^{++} X_a \ =\     \partial^{++} X_a   +  \,\Omega^{cb} \frac {\partial^2 {\cal L}^{+4}}{\partial f^{+a}\partial f^{+b}} \, X_c \,.
     \ee
     The constraint \p{harm-pseudoreal} implies  that 
  \be
   f^{+a}(t,u) \ =\ x^{ja} u^+_j + \ {\rm higher\ harmonic\ terms} 
    \ee
  with $x^{ja}$ satisfying the pseudoreality condition \p{pseudoreal-V}. As explained in the Appendix, this means that $x^{ja}$ can be traded for $4n$ real coordinates $x^M$.  
  
  To find the metric, we should resolve the equation \p{cons-comp} for $f^{+a}$, then the equation  \p{A-} (with the fermion terms suppressed) for the auxiliary field $A^{-a}$, substitute the solution into the action \p{action} and integrate over harmonics. Skipping the details (see Refs. \cite{FIS-nonlin,glasses} for pedagogical derivations and the proof that the  metric  thus derived is hyper-K\"ahler), we quote the result. The component bosonic Lagrangian can be presented in the form
     \be 
       L^{\rm bos} \ =\ \frac 12 g_{jc, kb} \,   {\dot x}^{jc}  {\dot x}^{kb} \,,
        \ee
        where 
        \be
        \lb{metr-HK}
        g_{jc, kb}   = e^{\u{ja}}_{jc}  \, e^{\u{kb}}_{kb} \, \varepsilon_{\u{jk}} \, \Omega_{\u{ab} } 
       \ee
       In this expression, the ordinary indices are the symplectic world indices, while the underlined indices refer to the tangent space.
       The vielbeins $e^{\u{kb}}_{kb}$ are determined from the relation
        \be
        \lb{vielbein-def}
        M^{\u{b}}{}_a (\partial_{kb} f^{+a} ) \ =\ e^{\u{kb}}_{kb} u_{\u{k}}^+\, ,
         \ee
         where the {\it bridge} $M^{\u{b}}{}_a$ is determined as a solution to the equation
         \be
         \lb{eq-bridge}
         {\cal D}^{++}M^{\u{b}}{}_a  = \partial^{++} M^{\u{b}}{}_a + \Omega^{cb} \, \frac {\partial^2 {\cal L}^{+4}}{\partial f^{+a} \partial f^{+b}} M^{\u{b}}{}_c \ =\ 0 \, .
          \ee      
          {\it We reiterate:} 
          
          \begin{itemize}
          
          \item To determine the HK metric, given the prepotential ${\cal L}^{+4}$, we first find $f^{+a}$ as a solution to the equation \p{cons-comp}. 
          \item Then we solve the equation \p{eq-bridge} for the bridge. 
          In fact, this homogeneous equation has  infinitely 
    many solutions  interrelated by the transformation (a kind of gauge transformation)
      \be
      \lb{tau-gauge}
       M^{\u{b}}{}_a \ \to \ R^{\u{b}}{}_{\u{c}}  \, M^{\u{c}}{}_a 
        \ee
         with an arbitrary harmonic-independent $R^{\u{b}}{}_{\u{c}}$. And for an arbitrary $R$, the objects 
         $e^{\u{kb}}_{kb}$ defined in \p{vielbein-def} are not strictly speaking the vielbeins: the metric is not expressed via
         $e^{\u{kb}}_{kb}$ according to \p{metr-HK}, but its expression is more complicated. It is possible, however, to choose the gauge such that the expression \p{metr-HK} is valid. Note that the ``connection"
         $\Omega^{cb} \,  {\partial^2 {\cal L}^{+4}}/{\partial f^{+a} \partial f^{+b}}$ that enters \p{eq-bridge} belongs to the algebra $sp(2n)$. It is then possible to choose the solution for $M^{\u{b}}{}_a$ that belongs  to the group
         $Sp(2n)$. One can show that in this case the relation \p{metr-HK} holds.
         \item With the bridge in hand, we find the vielbeins using \p{vielbein-def} and then the metric. 
          \end{itemize}

This is a long way, but it works.           
  
  \section{A metric from Taub-NUT family}
  \setcounter{equation}0
   \subsection{Taub-NUT metric}
   To derive the Taub-NUT metric, we set $n=1$ and choose
   \be
\lb{L+4TN}
{\cal L}^{+4} \ =\ -\frac 1 2\, (q^{+1})^2 (q^{+2})^2 \, .
\ee
Having performed all the operations outlined above, we obtain
 \be
 \lb{metrx-TN}
ds^2 = \frac{2 +  x^1 \!\cdot\! x^2}{1+ x^1 \!\cdot \! x^2} \,
(x^2 \!\cdot \! dx^1  +  x^1 \!\cdot \! dx^2)^2 
   +\,4(1+  x^1 \!\cdot\! x^2) dx^1 \!\cdot\! dx^2\,, \qquad  \label{ds-TN1}
 \ee
where $  x^1 \!\cdot\! x^2 = x^{j1} x_{j2}$, etc.

To bring the metric \p{metrx-TN} to the form known in the literature, we introduce the variables 
\be
\lb{def-3X}
X^1 = x^{21} x^{22} - x^{11} x^{12},\ \  X^2 &=& i(x^{11} x^{12} + x^{21} x^{22}), \ \ 
X^3 = x^{11} x^{22} + x^{21} x^{12}, \nn
\Psi  &=&  i\,\ln \frac{x^{22}}{x^{11}}.
\ee
Then \p{metrx-TN} coincides up to the extra irrelevant factor 2 with 
\be
\lb{metrX-TN}
ds^2\ =\ V(r) \, d\vecg{X}d\vecg{X} +  V^{-1}(r)\,  \left(d\Psi+ \vecg{A}d\vecg{X}\right)^2 \,,
\ee
where $\Psi \in (0, 2\pi)$, 
\begin{equation}\label{V-TN-12}
V(r) \ = \ \frac{1}{r} + 1
\end{equation}
($r = |\vec{X}|$) and
\be
\label{U-TN}
A^1 \ =\  \frac{X^2}{r\left(r+X^3 \right)}\, , \ \ \
A^2 \ = \  -\frac{X^1}{r\left(r+X^3 \right)}\, , \ \ \
A^3 \ =\ 0
\ee
is the vector potential of a magnetic monopole. 
The metric \p{metrX-TN} has nontrivial physical applications. In particular, a similar metric, but with the inverse sign of the ``Coulomb  potential  term" $1/r$ in \p{V-TN-12} describes the dynamics of 
two interacting BPS magnetic monopoles  at the distances much larger than their size. \cite{Gib-Man}.\footnote{Such metric involves a singularity  at $r=1$. Physically, this singularity smoothes out: the dynamics of BPS monopoles at small distances is described by a more complicated Atiyah-Hitchin metric \cite{AH}, which is smooth and goes over to the TN metric in the limit of large $r$. These details are irrelevant for us, but we have chosen to mention them to give a reader a proper perspective.}

\subsection{8-dimensional calculation}
We take $n=2$ and choose
 \be 
 \lb{L+4TN}
 L^{+4} = -q^{+1} q^{+2} q^{+3}q^{+4}\,.
  \ee
  The equations \p{cons-comp} read
  
  \be
\partial^{++} f^{+1} \ =\ f^{+1} f^{+3} f^{+4}, \qquad \partial^{++} f^{+2} \ =\ - f^{+2} f^{+3} f^{+4},  \nn
\partial^{++} f^{+3} \ =\ f^{+3} f^{+1} f^{+2}, \qquad \partial^{++} f^{+4} \ =\ - f^{+4} f^{+1} f^{+2}.
 \ee
 
 Their solution is
 
 \be
 f^{+1} \ =\ x^{+1} e^{\cal K}, \quad  f^{+2} \ =\ x^{+2} e^{- {\cal K}},
 \quad  f^{+3} \ =\ x^{+3} e^{\cal J}, \quad  f^{+4} \ =\ x^{+4} e^{- {\cal J}},
  \ee
  where 
  \be
  {\cal J} \ =\ \frac 12 (x^{+1} x^{-2} + x^{-1} x^{+2} ), \qquad 
{\cal K} \ =\ \frac 12 (x^{+3} x^{-4} + x^{-3} x^{+4} ).
    \ee
    Our next task is to solve the equation \p{eq-bridge}. As was mentioned above, this equation has many solutions, and to ensure \p{metr-HK}, we have to pick up a solution that belongs to $Sp(4)$. However, this requirement does not rigidly fix the solution yet: one can still multiply it by a constant $Sp(4)$ matrix. It is convenient to choose the solution which reduces to the unit matrix when the nonlinearity associated with ${\cal L}^{+4}$ is switched off. 
    
    The bridge reads

     \be
   M^{\u{b}}{}_a  \ =\ \frac 1{\Delta^{1/4}} \left( \begin{array}{cccc}
     e^{-{\cal K}} & 0 & - x^{-1} x^{+4} e^{-{\cal J}} & - x^{-1} x^{+3} e^{\cal J} \\
     0 & e^{\cal K} & x^{-2} x^{+4} e^{-{\cal J}} & x^{-2} x^{+3} e^{\cal J} \\
     -x^{-3} x^{+2} e^{- {\cal K}} & - x^{-3} x^{+1} e^{\cal K} & e^{-{\cal J}} & 0 \\
     x^{-4} x^{+2} e^{- {\cal K}} & x^{-4} x^{+1} e^{{\cal K}} & 0 & e^{\cal J} 
      \end{array}
      \right),
     \ee
    where the index $\u{b}$ marks the lines, the index $a$ marks the columns,
    \be 
    \lb{Delta}
    \Delta = 1 - (x^1 \cdot x^2) (x^3 \cdot x^4),
    \ee
     and we used the notation $(a \cdot b) \equiv a^j b_j$.

     The derivatives of $f^{+a}$ that enter the equation  \p{vielbein-def} are:

     \be
     &&\partial_{k1} f^{+1} \ =\  u^+_k e^{\cal K}, \qquad 
     \partial_{k2} f^{+2} \ = \ u^+_k e^{-\cal K}, \qquad   \partial_{k1} f^{+2} =  \partial_{k2} f^{+1} \ =\  0\,, \nn 
      &&\partial_{k3} f^{+3} \ =  \ u^+_k e^{\cal J}, \qquad   \partial_{k4} f^{+4} \ = \ u^+_k e^{-\cal J}, \qquad 
       \partial_{k3} f^{+4} = \partial_{k4} f^{+3} \ =\  0\,, \nn
     &&\partial_{k1} f^{+3} = \frac 12 x^{+3} (u^+_k x^{-2} + u^-_k x^{+2}) e^{\cal J}, 
     \qquad \partial_{k1} f^{+4} = - \frac 12 x^{+4} (u^+_k x^{-2} + u^-_k x^{+2}) e^{-\cal J}, \nn
      &&\partial_{k2} f^{+3} = \frac 12 x^{+3} (u^+_k x^{-1} + u^-_k x^{+1}) e^{\cal J}, \qquad \partial_{k2} f^{+4} = -\frac 12 x^{+4} (u^+_k x^{-1} + u^-_k x^{+1}) e^{-\cal J}, \nn
      &&\partial_{k3} f^{+1} = \frac 12 x^{+1} (u^+_k x^{-4} + u^-_k x^{+4}) e^{\cal K}, \qquad \partial_{k3} f^{+2} = -\frac 12 x^{+2} (u^+_k x^{-4} + u^-_k x^{+4}) e^{-\cal K}, \nn 
      &&\partial_{k4} f^{+1} = \frac 12 x^{+1} (u^+_k x^{-3} + u^-_k x^{+3}) e^{\cal K},      
  \qquad
   \partial_{k4} f^{+2} = - \frac 12 x^{+2} (u^+_k x^{-3} + u^-_k x^{+3}) e^{-\cal K}\,.
   \ee
   
  This gives the following vielbeins:
   
   \be
   e^{\u{k1}}_{k1} &=&  e^{\u{k2}}_{k2} \ = \  e^{\u{k3}}_{k3} \ = \ e^{\u{k4}}_{k4} \ = \ \Delta^{-1/4} \delta^{\u{k}}_k\,, \nn 
   e^{\u{k1}}_{k2}&=&  0, \quad  e^{\u{k1}}_{k3} = \frac {\Delta^{-1/4}} 2 [\delta^{\u{k}}_k (x^1 \cdot x^4) - x^{\u{k}4} x^1_k],  \quad  e^{\u{k1}}_{k4} = \frac  {\Delta^{-1/4}}2 [\delta^{\u{k}}_k (x^1 \cdot x^3) - x^{\u{k}3} x^1_k], \nn
    e^{\u{k2}}_{k1} &=& 0, \quad  e^{\u{k}2}_{k3} = \frac {\Delta^{-1/4}} 2 [\delta^{\u{k}}_k (x^4 \cdot x^2)  + x^{\u{k}4} x^2_k],  \quad  \ e^{\u{k2}}_{k4} = \frac {\Delta^{-1/4}}2 [\delta^{\u{k}}_k (x^3 \cdot x^2) + x^{\u{k}3} x^2_k], \nn
     e^{\u{k3}}_{k4} &=& 0, \quad   e^{\u{k3}}_{k1} = \frac {\Delta^{-1/4}}2 [\delta^{\u{k}}_k (x^3 \cdot x^2) - x^{\u{k}2} x^3_k],  \quad  e^{\u{k3}}_{k2} = \frac {\Delta^{-1/4}}2 [\delta^{\u{k}}_k (x^3 \cdot x^1) -  x^{\u{k}1} x^3_k], \nn
     e^{\u{k4}}_{k3} &=& 0, \quad  e^{\u{k4}}_{k1} = \frac {\Delta^{-1/4}}2 [\delta^{\u{k}}_k (x^2 \cdot x^4) + x^{\u{k}2} x^4_k],  \quad  e^{\u{k4}}_{k2} = \frac {\Delta^{-1/4}}2 [\delta^{\u{k}}_k (x^1 \cdot x^4)] +  x^{\u{k}1} x^4_k]\,.
   \ee

  The components of the metric are

   \be
   \lb{metr}
  \Delta^{1/2} g_{j1, k1} &=& \frac  14 [(x^2 \cdot x^4) (23)_{jk} - \{3 \leftrightarrow 4\} ], \qquad
     \Delta^{1/2}g_{j2, k2} \ = \ \frac 1 4 [(x^1 \cdot x^4) (13)_{jk} - \{3 \leftrightarrow 4\} ], \nn
     \Delta^{1/2}g_{j3, k3} &=& \frac 1 4 [(x^1 \cdot x^4) (24)_{jk} - \{1 \leftrightarrow 2\} ], \qquad
     \Delta^{1/2}g_{j4, k4} \ =\ \frac  14 [(x^1 \cdot x^3) (23)_{jk} - \{1 \leftrightarrow 2\} ],\nn
     \Delta^{1/2}g_{j1, k3} 
     &=&  [\varepsilon_{jk} (x^4 \cdot x^2) + x^4_j x^2_k], \qquad
     \Delta^{1/2} g_{j1, k4} \ = \ [\varepsilon_{jk} (x^3 \cdot x^2) + x^3_j x^2_k], \nn 
      \Delta^{1/2}g_{j2, k3} &=& [\varepsilon_{jk} (x^4 \cdot x^1) + x^4_j x^1_k], \qquad
      \Delta^{1/2}g_{j2, k4} \ =\ [\varepsilon_{jk} (x^3 \cdot x^1) + x^3_j x^1_k], \nn
      \Delta^{1/2}g_{j1, k2} &=&   \varepsilon_{jk} \left[ 1 + \frac 14 (x^2 \cdot x^4) (x^1 \cdot x^3)
     - \frac 14 (x^2 \cdot x^3) (x^1 \cdot x^4) \right]  \nn 
    && + \ \frac 14 \left[  (x^2 \cdot x^4) x^1_j x^3_k + (x^1 \cdot x^4) x^3_j x^2_k  
     - (x^1 \cdot x^3) x^4_j x^2_k - (x^2 \cdot x^3) x^1_j x^4_k\right]  \nn
    && + \ \frac 14 (x^1 \cdot x^2) (x^3_j x^4_k - x^4_j x^3_k), \nn
      \Delta^{1/2}g_{j3, k4} &=&   \varepsilon_{jk} \left[ 1 + \frac 14 (x^2 \cdot x^4) (x^1 \cdot x^3)
     - \frac 14 (x^2 \cdot x^3) (x^1 \cdot x^4) \right]  \nn 
     &&+ \ \frac 14 \left[  (x^1 \cdot x^4) x^3_j x^2_k + (x^1 \cdot x^3) x^2_j x^4_k  
     - (x^2 \cdot x^4) x^3_j x^1_k - (x^2 \cdot x^3) x^1_j x^4_k\right]  \nn
     &&+ \ \frac 14 (x^3 \cdot x^4) (x^1_j x^2_k - x^2_j x^1_k), 
     \ee
where $(23)_{jk} = x^2_j x^3_k + x^3_j x^2_k$, etc.

\subsection{Discussion}

If $x^a \gg 1$, the metric simplifies, acquiring a block-diagonal form. The metric components behave then as $g_{MN} \sim x^2$. The invariant volume 
$\int \! \sqrt{g} \, d^8x$ diverges.  The metric has thus similar properties to the Taub-NUT metric \p{metrx-TN} and represents an 8-dimensional generalization of the latter.

Three 8-dimensional generalizations of the Taub-NUT metric are known. They are related to the three classical simple Lie algebras of rank 2: $su(3), \ sp(4) = so(5)$, and $g_2$. The explicit expression for the 
$su(3)$  metric is \cite{N-monopoles}:
  \be
  \lb{su3}
  ds^2[su(3)] \ =\ \sum_{m,l = 1,2,3} A_{ml} \, d \vecg{X}_m d \vecg{X}_l \quad + \ {\rm phase\ part}\,,
   \ee 
   where $\vecg{X}_m$ are 3-dimensional vectors satisfying the constraint $\vecg{X}_1 + \vecg{X}_2 + \vecg{X}_3 = 0$,
   \be
   \lb{Amm}
   A_{mm} \ =\  1 + C \sum_{l \neq m} \frac 1{|\vecg{X}_m - \vecg{X}_l |} \qquad \qquad ({\rm no\ summation \ over\ } m)
   \ee
   and 
   \be
   \lb{Aml}
  A_{m \neq l} \ =\   - \frac C{|\vecg{X}_m - \vecg{X}_l|}\, ,
   \ee
  where $C$ is a real constant.
   
   This metric (with negative $C$ !) describes the dynamics of three well-separated BPS monopoles. Then $\vecg{X}_m$ are their positions.
  
  The explicit expressions for the $so(5)$ and $g_2$  metrics were found in \cite{Selivanov}. They are:
    \be
  \lb{so5}
  ds^2[so(5)] \ =\ \sum_{m=1,2}  d \vecg{X}_m^2  +  C\left[\sum_{\pm}  \frac
  {(d\vecg{X}_1 \pm d\vecg{X}_2)^2}{|\vecg{X}_1 \pm \vecg{X}_2|} + \sum_{m = 1,2} \frac {d\vecg{X}_m^2}{X_m}\right] + \ {\rm phase\ part}
   \ee
   and 
    \be
  \lb{g2}
  ds^2[g_2] \ =\ \sum_{m=1}^3 d \vecg{X}_m^2  +  C\left[\sum_{l<m\leq3}  \frac
  {(d\vecg{X}_m - d\vecg{X}_l)^2}{|\vecg{X}_m -  \vecg{X}_l|} + 3\sum_{m=1,2,3} \frac {d\vecg{X}_m^2}{X_m}\right] + \ {\rm phase\ part}
   \ee
 with the constraint $\vecg{X}_1 + \vecg{X}_2 + \vecg{X}_3 = 0$.
  The reader has recognized in the structures entering the right-hand sides of Eqs. \p{Amm} - \p{g2} the roots of the corresponding algebras. 
 
The metric \p{so5} can be obtained from the $su(4)$ metric describing the dynamics of four BPS monopoles by the hamiltonian hyper-K\"ahler reduction in the spirit of \cite{Hitchin,Rychenkova}. The metric \p{g2} can be obtained by a reduction of the  metric describing the dynamics of seven monopoles. (7 is the lowest dimension of a unitary group where $G_2$ can be embedded.)

A natural conjecture is that the newly derived metric \p{metr} is equivalent to one of these known hyper-K\"ahler metrics. If so, it is clear with  which one. The metric \p{metr} has the $Z_2$ isometry corresponding to the simultaneous interchange\footnote{It also  has  the continuous $SU(2)$ isometry associated with the rotations of the spinor indices $j,k$. This isometry, associated with the rotations of $\vecg{X}_m$, is present also in \p{so5} [and in \p{su3}, \p{g2}].} $x_1 \leftrightarrow x_3\, , x_2 \leftrightarrow x_4$. Also the metric \p{so5} has the $Z_2$ isometry associated with the interchange $\vecg{X}_1 \leftrightarrow \vecg{X}_2$. On the other hand, the metrics \p{su3} and \p{g2} have both the $S_3$ isometry  associated with the interchange of $\vecg{X}_{m=1,2,3}$, which the metric \p{metr} does not possess. (Obviously, these isometries are related to the Weyl groups of the corresponding algebras.) 

The reader may ask at this point: how come the isometry of the metric \p{metr} is only $Z_2$, while the harmonic prepotential \p{L+4TN}, from which this metric was derived was invariant under $S_4$ ? This apparent paradox is resolved by noting that we have chosen the particular convention \p{Omega} for the symplectic matrix $\Omega^{ab}$. The transformation $(12) \leftrightarrow (34)$ is
the only discrete transformation leaving this matrix invariant. 

Thus, we guess that the metric \p{metr} can be reduced to \p{so5} by a coordinate transformation expressing $\vecg{X}_m$ as a quadratic function of $x$---an analog of the transformation \p{def-3X}. It would be interesting to find this transformation explicitly and confirm this guess.

Given that the prepotential \p{L+4TN} gives the metric \p{so5}, a natural question is what prepotentials give the metrics \p{su3} and \p{g2}. Unfortunately, we cannot now answer this question. The procedure described in Sect. 2 is ``asymmetric": we can obtain the metric from the prepotential, but we  know no regular way to derive the prepotential from the metric. On the other hand, an exact mathematical proof of the fact that {\it any} HK metric can be derived from {\it some} prepotential ${\cal L}^{+4}(q^{+a})$ has been constructed \cite{L+4-HK-teor}.

  \section*{Appendix: Symplectic notation}
  \setcounter{equation}0
  
  \renewcommand\theequation{A.\arabic{equation}}
  
 We trade the vector index $M$ for a pair $(ja)$, where $j = 1,2$ and $a = 1, \ldots, 2n$.
  The indices are raised and lowered according to $X^i = \varepsilon^{ij}X_j, Y^a = \Omega^{ab} Y_b$,
 where $\varepsilon^{jk} = -\varepsilon_{jk}$ with the convention $\varepsilon_{12} = 1$ and 
$\Omega^{ab}$ is the symplectic matrix, which we choose in the form \p{Omega}.

 We introduce $4n$ rectangular matrices $\Sigma_M$,
\be
\label{Sigma}
(\Sigma_{1,2,3,4})^{ja} = \left( \sigma^\dagger_\mu\,, \, 0\,, \, \ldots \,,\, 0  \right)^{ja}, \  
(\Sigma_{5,6,7,8})^{ja}  = \left( 0\,,\, \sigma^\dagger_\mu\,, \, 0\,, \, \ldots \,,\, 0  \right)^{ja}, {\rm etc.}
\ee
with\footnote{And $(\sigma_\mu)_{aj} \  = \ \{(\vecg{\sigma})_{aj}, i\delta_{aj} \}$. Note the identities 
$$ \sigma_\mu \sigma_\nu^\dagger + \sigma_\nu \sigma_\mu^\dagger = 2 \delta_{\mu\nu}\,. $$}
\be
\lb{sigma-mu}
(\sigma^\dagger_\mu)^{ja} \  = \ \{(\vecg{\sigma})^{ja}, -i\delta^{ja} \}     \, . 
\ee
In \p{sigma-mu}, $a = 1,2$ and  $(\vecg{\sigma})^{ja}$ as well as $(\vecg{\sigma})_{aj}$ are the  {\it Pauli matrices}.

Then, for any tensor, we establish the correspondence
 \be
 \lb{sympl-vector}
T^{\ldots {ja} \ldots}  =  \frac i{\sqrt{2}} \, (\Sigma_M)^{ja}\, T^{\ldots M \ldots} \,, \quad
T^{\ldots M \ldots}  =  \frac i{\sqrt{2}} \, (\Sigma_M)_{ja}\, T^{\ldots ja \ldots}, \quad
 \ee
 where $(\Sigma_M)_{ja} = \varepsilon_{jk} \Omega_{ab} (\Sigma_M)^{kb}$, 
  and 
the dots stand for all other indices. Note that, for a real vector $V^M$, the components $V^{ja}$ obey the  {\it pseudoreality} condition
\be
\lb{pseudoreal-V}
\overline{V^{ja}} \ =\ \varepsilon_{jk} \Omega_{ab}  V^{kb} \equiv V_{ja} \, .
\ee

In these terms, the flat metric is expressed as
\be
\lb{metr-simpl}
g^{ja,\, kb} \ =\  -\frac 12\, (\Sigma_M)^{ja}  (\Sigma_M)^{kb}  =
\varepsilon^{jk} \Omega^{ab}  \, , \nn
g_{ja,\, kb} \ =\   \varepsilon_{jk} \Omega_{ab} \ = \ g^{ja,\, kb}\, .
\ee

The quaternionic triple of the tangent space complex structures can be chosen as
\be
\lb{Ip}
(I^p)_{ja, kb} \ =\  
-i(\sigma^p)_{jk} \, \Omega_{ab}\, , 
 \ee
 where 
 \be
\lb{symm-jk}
(\sigma^p)_{jk} \ =\ (\sigma^p)_{kj} \ =\ \varepsilon_{kl}
(\sigma^p)_{j}{}^{l}
  \ee
[$(\sigma^p)_{j}{}^{l}$ are the ordinary Pauli matrices].


\vspace{1mm}


 






\begin{thebibliography}{96}

 \bibitem{AG-F} L.~Alvarez-Gaum\'e and D.Z.~Freedman, {\it Geometrical structure and ultraviolet finiteness of the supersymmetric $\sigma$-model}, Commun. Math. Phys. {\bf 80} (1981) 443.

 \bibitem{Toppan} A.~Pashnev and  F.~Toppan, {\it On the classification
of $N$ extended supersymmetric quantum systems}, J. Math. Phys. {\bf 42} (2001) 5257, {\tt arXiv:hep-th/0010135}.

\bibitem{HSS}  A.S.~Galperin, E.A.~Ivanov, V.I.~Ogievetsky and E.S~Sokatchev, {\it Harmonic superspace}, Cambridge Univ. Press, 2001.

\bibitem{L+4-HK} A.S.~Galperin, E.A.~Ivanov, V.I.~Ogievetsky and E.S.~Sokatchev, {\it  Hyper-K\"ahler metrics  and harmonic superspace}, 
Commun. Math. Phys. {\bf 103} (1986) 515;

{\it Eguchi-Hanson type metric from  harmonic superspace}, Class. Quantum Grav. {\bf 3} (1986) 625.

\bibitem{DI}  F.~Delduc and E.~Ivanov, {\it ${\cal N} = 4$ mechanics of general ({\bf 4}, {\bf 4}, {\bf 0}) multiplets}, Nucl. Phys. {\bf B855} (2012) 815, {\tt arXiv:1104.1429 [hep-th]}

\bibitem{IL} E.A.~Ivanov and O.~Lechtenfeld, {\it ${\cal N}=4$ supersymmetric mechanics in harmonic  superspace}, JHEP {\bf 0309} (2003) 073,  {\tt arXiv:hep-th/0307111}.

\bibitem{glasses} A.V.~ Smilga, {\it Differential geometry through supersymmetric glasses}, World Scientific, 2020.

\bibitem{FIS-nonlin}  S.A.~Fedoruk, E.A.~Ivanov and A.V.~Smilga, {\it Generic HKT geometries in the harmonic superspace approach},  J. Math. Phys. {\bf 59}, 083501 (2018), {\tt arXiv:1802.09675 [hep-th]}.


\bibitem{Gib-Man} G.W. Gibbons and N.S. Manton, {\it Classical and Quantum Dynamics of BPS Monopoles},     Nucl.Phys. {\bf B274} (1986) 183.

\bibitem{AH} M. Atiyah and N. Hitchin, {\it The Geometry and Dynamics of Magnetic Monopoles}, Princeton University Press, 1988.

\bibitem{N-monopoles}
G.W. Gibbons and N.S. Manton,
{\it The Moduli space metric for well separated BPS monopoles}, Phys. Lett. {\bf 356B}, 32 (1995), 
{\tt arXiv: hep-th/9506052};\\
G. Chalmers and A. Hanany, {\it Three-dimensional gauge theories and monopoles}, 
 Nucl.Phys. {\bf B489} (1997) 223, {\tt arXiv:hep-th/9608105 [hep-th]}.

\bibitem{Selivanov} K.G. Selivanov and A.V. Smilga, {\it Effective Lagrangian for $3d\ N=4$ SYM theories for any gauge group and monopole moduli spaces},     JHEP {\bf 12}  (2003) 027, {\tt arXiv:hep-th/0301230 [hep-th]}.

\bibitem{Hitchin} N.J.~Hitchin, A.~Karlhede, U.~Lindstrom  and M.~Rocek, {\it Hyperkahler Metrics and Supersymmetry},  Commun. Math. Phys. {\bf 108} (1987) 535. 

\bibitem{Rychenkova} G.W.~Gibbons and P.~Rychenkova, {\it HyperKahler quotient construction of BPS monopole moduli spaces}, Commun. Math. Phys. {\bf 186} (1997) 585, {\tt arXiv:hep-th/9608085}.

   
\bibitem{L+4-HK-teor} A.S.~Galperin, E.A.~Ivanov, V.I.~Ogievetsky and E.S.~Sokatchev, {\it Gauge field geometry from complex and harmonic analyticities. II Hyper-K\"ahler case.}, Ann. Phys. {\bf 185} (1988) 22.







 

 
 
 
 


\end{thebibliography}
\end{document}